# SEQUENTIAL CATEGORY AGGREGATION AND PARTITIONING APPROACHES FOR MULTI-WAY CONTINGENCY TABLES BASED ON SURVEY AND CENSUS DATA[1]

By L. Fraser Jackson, Alistair G. Gray and Stephen E. Fienberg

*Victoria University of Wellington, Statistical Research Associates and Carnegie Mellon University*

Large contingency tables arise in many contexts but especially in the collection of survey and census data by government statistical agencies. Because the vast majority of the variables in this context have a large number of categories, agencies and users need a systematic way of constructing tables which are summaries of such contingency tables. We propose such an approach in this paper by finding members of a class of restricted log-linear models which maximize the likelihood of the data and use this to find a parsimonious means of representing the table. In contrast with more standard approaches for model search in hierarchical log-linear models (HLLM), our procedure systematically reduces the number of categories of the variables. Through a series of examples, we illustrate the extent to which it can preserve the interaction structure found with HLLMs and be used as a data simplification procedure prior to HLL modeling. A feature of the procedure is that it can easily be applied to many tables with millions of cells, providing a new way of summarizing large data sets in many disciplines. The focus is on information and description rather than statistical testing. The procedure may treat each variable in the table in different ways, preserving full detail, treating it as fully nominal, or preserving ordinality.

Received February 2007; revised November 2007.

[1]Supported in part by a research grant from the Official Statistics Research and Data Archive Center, Statistics New Zealand and by the National Science Foundation under Grant DMS-06-31589 to the Department of Statistics, Carnegie Mellon University and Grant DMS-04-39734 to the Institute for Mathematics and Its Application at the University of Minnesota. The ideas expressed here are the sole responsibility of the authors and are not associated with any official position or policies within Statistics New Zealand.

**1. Introduction.** Finding parsimonious summaries of data sets and contingency tables generated from them has been a long term objective of statisticians. The number of cells in a contingency table depends on the number of variables $K$, and the number of categories for each of the variables $r_k$ ($k = 0, \ldots, K-1$). These numbers can quickly become large. The traditional method of analysis using hierarchical log-linear models (HLLMs) as described in Bishop, Fienberg and Holland (1975), Fienberg (1980), Haberman (1978) and Whittaker (1990) constructs an analysis using the interaction terms between the variables. There are $2^K$ interaction terms, and the modeling process seeks to eliminate any terms which are not necessary for describing the interactions in the data.

For small tables this process can work well, but for large sparse data of the sort arising in survey and census contexts this process presents many difficulties. The number of cells $\prod r_k$ is not less than $2^K$ and indicates that the number of parameters required grows much more rapidly than the number of interaction terms. To model the two factor interaction for variables $i$ and $j$ requires $(r_i - 1)(r_j - 1)$ terms. So a summary constructed using two factor interaction terms may still involve large numbers of parameters and high order interactions, involving even larger numbers.

These difficulties are compounded by experience with large data sets such as national censuses conducted by official statistical agencies. In tables for such data sets the number of observations may be so large that on any conventional criteria it may be impossible to reject the saturated model. In attempting to construct an HLLM of the table we find we cannot eliminate any interaction terms and the complete table is the only *valid* summary. In some ways that is a useful result, but if the table contains a million or more cells, reporting the full table is typically an inadequate way of conveying its structure in terms easily understood and interpreted by human users. Both the significance of the interaction terms and the magnitude and signs of their parameters are important. Sparsity can and does still arise in such settings, thus collapsing makes even more sense since the presence of large numbers of sampling zeros can often lead to the nonexistence of maximum likelihood estimates or to minimal information about various parameter estimates, for example, see the discussion in Fienberg and Rinaldo (2007) and Dobra et al. (2008).

In this paper we approach the problem of constructing a summary of a table by examining the category structure within the variables and reducing the number of categories $r_k$. If the behavior of two categories is sufficiently similar, we collapse them by aggregating the counts in the cells and forming a new combined category. This leaves the set of variables unchanged, but aims to simplify the manner in which their interactions are described. It can be thought of as examining the saturated model between all pairs of categories and modeling the pair associated with least information loss as having the same behavior. We will use the terms aggregating two categories and collapsing two categories as synonyms and will refer to it as Paired Category Collapsing (PCC). In this process the number of variables is unchanged, but the size of the table necessary to model it is rapidly reduced. We explore this process and its properties and show in many cases it provides a range of insights into the structure of the associations in the table. For the smaller tables obtained we recommend fitting an HLLM to examine the interaction structure of the variables.

In summary, an analysis can be thought of as having two steps. First, consider using a principled way of reducing the number of cells without necessarily reducing the number of variables (the PCC steps). Second, on examination of the results of that process select and fit an HLLM to the data. We refer to a model fitted to the category partitions derived with PCC steps as a hierarchical log-linear partition model (HLLPM). This paper focuses on the PCC steps since the second stage is already well known, but we also suggest a modification of the criteria in the HLLM analysis.

Official statistical agencies have other reasons for wanting to use the PCC process. In large tables the most frequent cell counts are commonly "0," "1," or "2." But legislated requirements often preclude release of identifiable information about respondents. In a table both "1"s and "2"s can contribute to revealing such information, and thus agencies must take steps to publish their data in ways which prevent this. The PCC process provides a systematic method of simplifying a table to assist in satisfying these requirements. The simplification is often called *collapsing* or *coarsening* in the statistical literature [e.g., see Lauritzen (1996) and the references contained therein] and *global recoding* in the confidentiality literature [e.g., see Willenborg and de Waal (2000) who use the concept of minimizing information loss to consider alternatives]. This paper also applies the concept of minimizing information loss in a systematic way, and links it to a class of models and their maximum likelihood estimators. For nearly all tables we have studied, this class provides models which perform better than the usual sequence of HLLMs if a small number of parameters to describe the data is an important requirement of the summary.

The PCC process constructs a partition or coarsening of the categories for each variable. Such coarsening is typically not *coarsening at random* [cf., Heitjan and Rubin (1991) and Jaeger (2005)] and thus there is a loss of information relative to the original sets of categories. Papers by Lancaster (1949, 1951), Goodman (1968, 1970), Kreiner (2003) and many others have explored alternative methods of partitioning the data in tables mainly for significance testing. Gokhale and Kullback (1978) provide illustrations, applications and extensions of theses ideas. Important results of this literature are summarized in Gilula and Haberman (1998). Our procedure is an application of a class of models they discuss. Goodman (1981) examined the problem of simplifying the description of social class inheritance of persons in various groups, but his procedures do not generalize easily. Wermuth and Cox (1998) examine ways of exploring relationships between categories in ordinal systems and make extensive use of local odds ratios to construct summaries. We examine a table they studied below. Dellaportas and Tarantola (2005) outlined a Bayesian approach to jointly exploring category partitions and model structure. Their procedures are only feasible for tables of modest size. The class of level hierarchic models they define is a standard HLLM applied to a partition model. Kreiner (2003) includes procedures for combining categories in his programs but has a different focus from here.

In Section 2 we outline the information measures we use and the sequential category aggregation procedure. We show that it is constructing data consistent with a maximum likelihood model with some constraints across its parameters. Successive steps reduce the total number of parameters in the model. In Section 3 we discuss briefly the relationship between our procedures and the extensive collapsibility literature. In Section 4 we examine a simple table used by Wermuth and Cox and illustrate how using information loss associated with a collapsing step adds insights to analysis of even two way tables. In Section 4.3 we examine a four way table used by Christensen (1997), and in Section 5 explore application of these ideas to very large tables. Section 6 provides some concluding comments. Programs are provided in Jackson, Gray and Fienberg (2008b).

## 2. Data structure and information measures.

2.1. *Notation and information measure.* We will assume that the data consist of observations on a set of $K$ variables, $\{x_0, x_2, \ldots, x_{K-1}\}$, with $\{r_0, r_1, r_2, \ldots, r_{K-1}\}$ categories, respectively. Thus, we organize the data into a contingency table $N$ with cell counts $n_i$, where $i$ is an index ranging over the product set of the integers, $r$. The total number of observations is $n = \sum_i n_i$, the table shape is given by the vector of category numbers, and number of cells by $r = \prod_k r_k$.

The family of hierarchical log-linear models (HLLMs) implicitly provides a parameterization for such a table by using the set of all marginal tables of the $K$-dimensional array. Each marginal table $m$ corresponds to an additional $d_m = \prod_{k \in m}(r_k - 1)$ parameters. The set $M$ of marginal tables consists of $2^K$ members. The empty set is a member, and represents the model with the same frequency for all cells. We order the elements of $M$ using a binary representation with the dimensions taken in sequential order.

To provide a simple link between these definitions and the common representation using variables $A, B, C, \ldots,$ use the expressions $A$, $B$, $AB$, $C$, $AC$, $BC$, $ABC$,... to represent the sequence of marginal tables, and $(\mu)$, $\mu_A$, $\mu_B$, $\mu_{AB}$, $\mu_C$, $\mu_{AC}$, $\mu_{BC}$, $\mu_{ABC}$,... to represent the corresponding set of terms in the log-linear model with

$$\log(E(n_i)) = \sum_{m \in M} \mu_m(i).$$

We use the multinomial model for the observed cell counts, where $\boldsymbol{\pi} = \{\pi_i\}$ are the cell probability under the model and $\mathbf{p} = \{p_i = n_i/n\}$ are the observed cell proportions. The kernel of the log likelihood function $L(\boldsymbol{\pi}|\mathbf{n})$ has the form

$$\sum_i n_i \log(n\pi_i) = n \sum_i p_i \log(n\pi_i).$$

One way to assess the fit of a specific HLLM is via the likelihood-ratio statistic

$$G^2 = 2n \sum_i p_i \log(np_i) - 2n \sum_i p_i \log(n\hat{\pi}_i),$$

which compares the fit of that model to the *saturated* model that includes all possible parameters and fits the data perfectly. When $\hat{\boldsymbol{\pi}}$, the estimates of $\boldsymbol{\pi}$, are computed via maximum likelihood, $G^2$ has an asymptotic $\chi_h^2$ distribution. In a complete table $h$ is the number of cells less the number of independent parameters in the model. This statistic reduces to

$$2n \sum_i p_i \log\left(\frac{p_i}{\hat{\pi}_i}\right),$$

which is just $2n$ times the Kullback–Leibler distance between $\mathbf{p}$ and $\hat{\boldsymbol{\pi}}$. The KL-distance is a measure of the information loss resulting from using the table with the expected values under the simpler model compared with the original (and saturated) table. We calculate the loss using a guarded method so $p \log(p)$ is zero for $p = 0$. Goodman (1971) gave heuristic grounds for using the gradient of the information loss $G^2/h$ in the model selection process and we discuss this later in Section 2.4.

In the analysis of contingency tables, interpretability of the model, particularly for high dimensions, is not easy unless the model is from the class of *independence* models. Thinking of a three-dimensional table, the *independence* models are *mutual independence*, with marginal tables $A, B, C$; *joint independence*, with, for example, marginal tables $AB, C$; *conditional independence* with, for example, marginal tables $AC, BC$. We can think of $C$ jointly independent of $A$ and $B$, as $C$ mutually independent of a supervariable $AB$. Interpretabilty is one motivation for our choice of models in the PCC procedure in the next section.

2.2. *Partitions and category aggregation.* A typical HLLM approach to summarizing a table is via the minimal sufficient statistics which are the marginals corresponding to the highest-order terms in the model. Thus, we can think of our goal as one of finding an adequate subset of the marginal tables. This process may drop certain of the original variables, but always preserves the categories in any variables present in the finally chosen set of marginal tables. As we mentioned in the introduction, our approach to summarizing a table involves examining the category structure within the variables and reducing the number of categories $r_k$. At any step, this leaves the set of variables unchanged, but aims to simplify the manner in which their interactions are described. Then, of course, we may also reduce the number of variables.

We denote by $u$ and $v$ categories of a variable $t$ which we consider for aggregation in the current table. The categories provide a two-category subset or collapsing of the source table, and we use $N_{(u,v),t}$ to represent this subtable and $N_{w,t}$ to represent all remaining cells. Within the subtable $N_{(u,v),t}$, the category aggregation is a variable collapsing operation and we can consider the two marginal tables $N_{u,t}$ and $N_{v,t}$. Conditions for its validity are discussed in Section 3. If these two tables have the same structure, then we have joint independence in $N_{(u,v),t}$ and no information about the interaction structure is lost in aggregation (although we may still not have coarsening at random). Otherwise, there is a loss of information about the interaction structure in the original table as we move to the new aggregated table. In providing summaries of the information in $N$, it is common to publish a marginal distribution for each possible $t$, so to avoid biasing the information loss, we always assess the loss as the Kullback–Leibler distance between $N_{(u,v),t}$ and the model of independence between the marginal table formed by summing across $u$ and $v$, and the totals for category $u$ and for category $v$.

Our PCC algorithm for a contingency table $N$ is as follows:

1. For each dimension $k$, construct a list of all possible pairwise category aggregations $(i, j)$.
2. For each aggregation above, compute $G^2$ and its degrees of freedom $h$ for the table formed from the two subtables with cells with category $i$ and category $j$ on dimension $k$.
3. Compute the quotients $G^2/h$.
4. Compute the dimension and category pair with the minimal quotient.
5. Aggregate the category pair, and return the new array.
6. Repeat from step 1 until the array is reduced to a single element in all but one dimension.

Any partition of the categories of a variable provides a possible method of aggregating them. For variable $k$ with $r_k$ categories in a $K$-dimensional table a recode is described by two components, the dimension $k$ to which the recode applies, and a vector of $r_k$ elements each a distinct element of which represents a group to be formed. An example for $r_k = 5$ is a vector of keys given by 2 2 3 2 5. This indicates the first, second and fourth categories are the first partition and categories 3 and 5 are a further two partitions. For simplicity, we adopt the convention that the new categories are in partitions in the order of first occurrence within the key vector. The number of partitions for one variable is reduced with each step. During the PCC process, the current table is formed by summing all categories in each partition.

By using this convention, each successive aggregation step provides an independent component in a partition of the total distance between the table and the model of marginal independence for all variables. The procedure permits exploration of a wide class of models where the full HLLM structure is retained, but within all marginal tables there is structure in the model probabilities. Within each term $\mu_m(i)$ all terms with indices from the same partition of included variables must have the same values.

Suppose we collapse cases in categories $i$ to a new set of combined categories $j$. Then the likelihood of the data given the new model is

$$L(\boldsymbol{\pi}'|\mathbf{n}) \propto \sum_j p_j' \log(p_j'/n\pi_j').$$

The difference between $L(\boldsymbol{\pi}|\mathbf{n})$ and $L(\boldsymbol{\pi}'|\mathbf{n})$ is a measure of information loss, but we need to be careful when calculating it. To calculate the information loss, we use the probability structure of the collapsed table, but expand each cell to the set of all cells in the same partitions of the original variables. We constructed the expansion to match the marginal distributions of the original table. Thus, we calculate the information loss using these expected values for all cells in the original table. Our analysis uses a decomposition of the likelihood ratio statistic (deviance) discussed by Gilula and Haberman (1998). As such, we can think of the information loss as either the total loss associated with a model of the full table, or as the sum of the information loss for all steps toward a smaller table.

2.3. *Reference models.* The graphical models literature emphasizes that variables in contingency tables may have very different roles. If an experiment is designed over a space of two variables $[AB]$, the design may impose some interaction structure which is not relevant in the comparisons made in the collapsing procedure. The analyst may not want to collapse over these categories, or study independently their effects.

If we exclude variables $A$ and $B$ from the collapsing process, then we always include the set of interactions between them in the model during the backward fitting process. It may still be of interest to consider what happens when we examine partitions of variables $A$ and $B$. To do so, we can exclude the remaining variables from the fitting process. That may suggest that aggregating over some of the design points will improve the efficiency of measuring effects.

In the description of our procedure we assumed that all variables are nominal. When the categories have an ordinal structure, there may be strong grounds for preserving it in any newly constructed categories. Then we consider forming pairs for only adjacent categories and this simplifies and speeds up the calculations.

All of these options are provided for by allowing a different treatment for each variable in the table. We have found that the three options, omitting a variable from the collapsing process, including it as an ordinal variable, or treating it as a nominal variable with all aggregates possible as in our examples, are all useful and essential for covering common analysis problems.

We may require further options for data on geographic units, especially when there are large numbers of such units and the $G^2$-statistics indicate significant differences but have many very close values. For such cases an alternative providing a clustering procedure may be appropriate, and geographic proximity may also be an important attribute to be considered in forming groups. Similar issues will arise with many social, economic, and health related classification systems which have a hierarchical structure and large numbers of categories.

2.4. *Selecting categories to collapse.* A recoding strategy must define a sequential process that will construct successive recodes. A single step approach involves considering the entire set of possible recodes, of which there are $\binom{r_k}{2}$ for each variable. For each such case, we construct the recode, expand the model to the original shape, and then calculate the information loss. Our algorithm processes the combinations in a lexical order, but that is not essential.

The selection of a recode raises a number of issues. The recodes for different variables may be associated with very large differences in the number of terms. To minimize the rate of information loss, we consider the gradient of information loss with respect to the number of estimated parameters. We select recodes at each step which minimize the mean information loss per parameter eliminated. This is different from the usual statistical procedure of using the significance level of the $G^2$ statistic and making a decision on the basis of the largest $p$-value computed by reference to the asymptotic $\chi^2$ distribution.

We have not followed that procedure here for two reasons. First, it seems more appropriate to focus on the information loss directly. If there are differences in the data structure for the marginal tables being compared, the asymptotic distribution of $G^2$ will be a noncentral $\chi^2$ with noncentrality parameter $\lambda$ giving a measure of the magnitude of information loss. The expected value of $G^2$ for degrees of freedom $r$ is $(\lambda + r)$ and the expected gradient is therefore $1 + \lambda/r$. Essentially, we use the gradient as an indicator of the magnitude of the information loss with a reduction of model complexity. Second, for the typical data sets of official statisticians there are very large numbers of cases $(n > 10^6)$ and anything other than the saturated model is usually rejected in tables of few dimensions. For many terms the unknown noncentrality is large and all terms may have very small $p$-values under the central $\chi$-squared distribution. The $p$-values may be difficult to calculate accurately and selecting on the basis of $p$ value becomes quite impractical. Further, for these tables they exhibit the characteristics of a large number of rare event models explored by Khamaladze (1988) and many others and the asymptotic distribution is no longer $\chi^2$, so decisions based on that distribution are inappropriate. For data sets based on small or moderate numbers of observations, the procedure could be modified to select on the conventional basis, however, it seems to work effectively even in those cases.

After each step, we use the new reduced array to consider a further step. If the optimal step is to collapse a variable to a single category, that is, essentially to aggregate over all categories, then we take that step. The process will terminate after at most $(\sum_k r_k) - (K+1)$ steps, at which stage the array will have only one category in each of its $K$ dimensions. It can terminate earlier if the table has block diagonal or triangular diagonal structure, or can be row permuted to either of these forms. There is no further change in our measure of information loss when the data are reduced to a single marginal vector.

While some readers might consider this procedure as just data dredging, we would reject that claim. First, we note that while a range of considerations may have led to the categories used in the data collection, the categories used to report the data are often ad hoc and we should not assume that the analyst got it right. Thus, we are letting the data determine ways in which we can reduce the size of the data space with limited loss of information for a given modeling framework.

Clearly, we could consider other recoding strategies. We have examined selection using the set of $\binom{r_k}{3}$ category triples, but they do not seem to have the flexibility we obtained using a single step aggregation and they increase the complexity of the computation rapidly. We could also use strategies based on category size, but the procedure we have adopted adjusts to size, and frequently ends up combining one of the smallest categories with another category.

A case can be made for using some form of penalized likelihood function, such as that associated with

$$\text{AIC} = -2 \times \log \text{likelihood} + 2 \times \text{number of parameters,}$$

or

$$\text{BIC} = -2 \times \log(\text{likelihood}) + \text{number of parameter} \times \log(\text{sample size}),$$

instead of the KL distance in the model selection stage of the process. We note that this does not alter the ordering for any pairs within a dimension. It may modify the selection of the variable for a collapse. Such choices will yield differences in cases where there are two or more pairs with small KL values and differences in their degrees of freedom. The differences in the KL values in many cases are large enough to offset any differences arising from making the AIC adjustment. The use of AIC or BIC would move us away from the concept of successive partitioning of the information, and thus, we have not pursued it further.

**3. The collapsibility literature and Simpson's paradox.** We can trace the collapsibility literature back at least to Yule (1903), and much later Simpson (1951) examining the case where aggregating groups of different size could lead to an apparent reversal of an effect. Bishop, Fienberg and Holland [(1975), pages 38–42] examined conditions for two-factor terms in a log-linear model of a three variable table being estimated consistently when collapsing over one of the variables. Whittemore (1978) clarified necessary and sufficient conditions, but see the discussion in Fienberg (1980). The notion of collapsing was extended and generalized by Asmussen and Edwards (1983) and others and is well summarized in Lauritzen's book Lauritzen (1996). Theorem 2 from Asmussen and Edwards shows that you can validly collapse over a variable whenever the saturated model on the remaining variables is a generator in the data generating process for the whole table. These results provide a framework within which our procedure can be justified if the saturated model on the remaining variables after the collapsing operation is a generator in the model of the source table. For the application to Census tables, this is generally a plausible condition.

The collapsibility literature poses a real problem for the statistical user. Generally, one's objective in collapsing over variables is to provide some simplification of a data set of unknown structure, while the conditions for its validity depend on knowledge of that structure. Hence, our approach has been to assume that there are components of the model associated with all terms in the hierarchical model, but to endeavor to find simplifications which will minimize the noncentrality parameter arising from aggregating categories. The strict conditions for collapsibility of categories depend on marginal or joint independence. Those conditions occur for a relatively small proportion of cases in our experience. It is nevertheless useful to explore what happens when they are not met, and departures from them are small relative to other components in the model.

It is clearly possible in a general collapsing process without restrictions to generate examples of Simpson's Paradox. In Jackson, Gray and Fienberg (2008a) we show that the PCC procedure throws an interesting light on conditions under which Simpson's paradox arises. While completely collapsing the variables of a table may generate such cases, in general they cause large losses of information and are avoided in the collapsing sequence.

## 4. Examples for two and four dimensional tables.

4.1. *Constructing a sequence of models of a table.* Wermuth and Cox (1998) give a table on schooling and age group from a social survey of adults in West Germany. The first dimension (0) corresponds to the rows and is a sequence of education categories and the second dimension (1) to the columns, a set of age groups. Both the variables in this table are ordinal variables. We note that all sequences begin with 0 in the tables below.

In their analysis of the data, Wermuth and Cox used identification of patterns in the standardized local odds ratios and their deviations from unity for pairs of cell entries. Consider an arbitrary pair of rows. If the odds ratios are the same for all columns, then in the table formed from that pair of rows, the entries display independence between the distribution across the rows and the distribution across the columns. $G^2$ provides a test of the similarity of the odds ratios for this pair of rows. Table 2 gives this distance for each possible pair of rows.

Having a formal summary statistic of the distances between all category pairs gives insight into the category structure. For the schooling categories in Table 2, category 1 is obviously very different from all the others except category 0. Combining categories 0 and 1 will simplify the category structure with little or no loss. The numbers are revealing about the confidence we might wish to assign to there being a difference between each pair of categories and highlight the large differences between category 1 and the remaining categories. If the data are from the range of HLLMs discussed by Birch (1963), then, if there is no underlying difference between the categories being compared, the distance measure will be distributed as chi-squared with 4 degrees of freedom. The table shows that nearly all possible aggregates lead to some information loss.

TABLE 1
*Observed counts in 1991/92 for schooling and age groups*

| | | Age | | | | |
|---|---|---|---|---|---|---|
| | | 18–29 | 30–44 | 45–59 | 60–74 | 75+ |
| **Schooling** | | **0** | **1** | **2** | **3** | **4** |
| Basic, incomplete | 0 | 12 | 13 | 12 | 20 | 7 |
| Basic, complete | 1 | 215 | 507 | 493 | 460 | 137 |
| Medium | 2 | 277 | 300 | 192 | 126 | 38 |
| Upper medium | 3 | 52 | 91 | 47 | 15 | 6 |
| Intensive | 4 | 233 | 225 | 102 | 74 | 19 |

TABLE 2
*Information loss for aggregation of row pairs—schooling categories*

|   | **0** | **1** | **2** | **3** | **4** |
|---|---|---|---|---|---|
| 0 |   | 6.95 | 20.44 | 32.92 | 30.40 |
| 1 |   |   | 173.69 | 77.52 | 236.06 |
| 2 |   |   |   | 14.77 | 12.99 |
| 3 |   |   |   |   | 16.31 |
| 4 |   |   |   |   |   |

To consider odds ratios in the columns rather than the rows, we construct Table 3. Those at the youngest age have the largest difference in educational profile from all other groups and those groups at 45 and above are much closer to each other than the others, even though there are significant differences. Combining the information in the two tables, we see that we can combine the two highest age categories without significant loss of information. Having done so, we recalculate the values in Table 2 and we find that the smallest entry is now 6.82 and in the same location as in the original table. Thus, we can combine the lowest two educational categories without any significant loss of information.

If we repeat the process outlined above until there are only cells in a single row, we obtain Table 4 which provides information about each of the sequential steps pairing two categories.

The columns of the table give us, respectively, a row identifier, the dimension on which the collapsing step takes place, the key vector for the current step expressed in terms of the original categories, the dimensions of the new table, the total information loss relative to the saturated model with the degrees of freedom for the current model and for the residuals, the change of deviance and its degrees of freedom to assess the current step, and the adjusted $R^2$ defined as $1 - \mathrm{dev}(r)\,\mathrm{dfres}(r=8)/(\mathrm{dev}(r=8)\,\mathrm{dfres}(r))$ for this table, and $r$ the current row.

TABLE 3
*Information loss for aggregation of column pairs—age categories*

|   | **0** | **1** | **2** | **3** | **4** |
|---|---|---|---|---|---|
| 0 |   | 70.52 | 178.53 | 253.15 | 117.20 |
| 1 |   |   | 43.25 | 110.11 | 45.81 |
| 2 |   |   |   | 23.96 | 10.13 |
| 3 |   |   |   |   | 0.84 |
| 4 |   |   |   |   |   |

TABLE 4
*Summary of sequence of PCC steps and models*

| $r$ | $d$ | | Key | | | | dim | | dev | dfmod | dfres | dev (term) | df | AdRsq |
|---|---|---|---|---|---|---|---|---|---|---|---|---|---|---|
| 0 | | | | | | | 5 | 5 | 0.00 | 24 | 0 | 0.00 | 0 | 1.000 |
| 1 | 1 | 0 | 1 | 2 | 3 | 3 | 5 | 4 | 0.84 | 20 | 4 | 0.84 | 4 | 0.991 |
| 2 | 0 | 0 | 0 | 1 | 2 | 3 | 4 | 4 | 7.66 | 17 | 7 | 6.82 | 3 | 0.951 |
| 3 | 0 | 0 | 0 | 1 | 2 | 1 | 3 | 4 | 20.39 | 14 | 10 | 12.73 | 3 | 0.909 |
| 4 | 0 | 0 | 0 | 1 | 1 | 1 | 2 | 4 | 35.69 | 11 | 13 | 15.30 | 3 | 0.877 |
| 5 | 1 | 0 | 1 | 2 | 2 | 2 | 2 | 3 | 52.89 | 10 | 14 | 17.20 | 1 | 0.831 |
| 6 | 1 | 0 | 0 | 1 | 1 | 1 | 2 | 2 | 110.54 | 9 | 15 | 57.65 | 1 | 0.670 |
| 7 | 0 | 0 | 0 | 0 | 0 | 0 | 1 | 2 | 357.15 | 8 | 16 | 246.61 | 1 | 0.000 |
| 8 | 0 | 0 | 0 | 0 | 0 | 0 | 1 | 2 | 357.15 | 8 | 16 | 0.00 | 1 | 0.000 |

Using conventional criteria in rows 1 and 2, the information loss is not significant. In row 3 a further step to reduce the education classification leads to a small but significant information loss (12.7 with 3 df), and row 4 gives a further step showing that the table can be reduced to a $2 \times 4$ table with a total information loss of 35.69 or approximately 10% of the information about the interaction between the categories on these two variables.

Even more striking is row 6. This shows that we can collapse this data to a single parameter model for the interaction between the variables with merely two categories for each variable and retain two thirds of the information in the data about the association between age categories and education levels. We can refine the model by refining the age classification as in line 5 and retain 83 percent of the information with a simple 2-parameter model. Note how we have identified more sensitivity to changes in the age structure than the educational categories.

In Table 4 the components of information are relative to the model of independence of the two variables. Distinguishing between basic education and other categories and using the four age groups formed by aggregating the highest two age categories preserves a high proportion of the information about the association between the two variables. This type of summary goes beyond the inferences of Wermuth and Cox, and is a member of a simple sequence of summaries with nested parameter sets.

In Table 5 we fit a HLLM to the data. Since we retain all main effects for the $5 \times 5$ table, there are two lines, one specifying the saturated model which has zero information loss, and the other based on the two main effects. The information in the line listed as main effects [A] [S] contains deviance between the saturated model and the marginal independence model, the degrees of freedom for the model and the residuals. Hence, there are only two points at which it gives the information loss. The collapsing process gives a systematic sequence of steps adding structure to the model and reducing the number of parameters required.

TABLE 5
*Fitted HLLM components*

| 2 Vars | | Shape 5 5 | | | | |
|---|---|---|---|---|---|---|
| Terms | Deviance | dfmod | dfres | dev (term) | df | Rsq |
| SA | 0 | 24 | 0 | 0 | 0 | 1 |
| AS | 357.146 | 8 | 16 | 357.146 | 16 | 0 |

Figure 1 provides a useful graphical display of the way in which the information loss is associated with the number of model parameters for the Wermuth and Cox data. In all cases the PCC model is expanded out to the original size array, and the original marginals are matched. It plots the loss of information with additional terms by model types, PCC and HLLM.

In the two variable case with a single interaction term [SA], there are only two points on the HLLM graph, corresponding to the saturated and independence modes. The blue line shows how information loss with PCC in fact tracks quite slowly upward initially and a small number of parameters provide most of the information. In Section 4.2 we find those parameters by looking at the marginal tables associated with the lines of the PCC summary.

While we express the sequence in terms of reducing the size of the data array, this is simply a tool for constructing models for the source table. The entries in Table 6 derive from the calculation of sufficient statistics at each step. The ratios in Table 7 show the structure of the coefficients for the model at each step of the process, and provide a direct comparison with the saturated model.

4.2. *Examining the departures from independence.* Goodman (1996) summarized the structure of several measures of association in a contingency table. A central component of all these measures is what he calls the $\psi$-statistics or *Pearson ratios*. For the two-dimensional table they are simply the ratio of the observed probability to the estimated probability under the independence model. Such ratios played a fundamental role in the original work of Lancaster (1951) and Kullback (1959) and also appear throughout the literature on correspondence analysis. Table 6 lists the summary tables for the collapsing process corresponding to each line of Table 4 and Pearson ratios for each summary table. It provides a picture of the way that the steps have simplified the interaction pattern between the variables. Table 7 gives these ratios for the model of the whole data set associated with each row of Table 4. The ratios characterize the summary of the table at each stage and enable a comparison with the saturated model. The log of the ratios between the matrices in row 0 and row $r$ give the residuals for the log-linear model fitted by the PCC procedure. Note that the numbers in the original table are important in assessing the significance of the residuals and they will have very different variances.

If we start with row 0 from Table 4, we have a saturated model and ratio of cell entries to the independence model. Later rows give the ratios after the successive collapsing steps. Direct comparison using Table 7 makes it easy to compare the observed model ratios with the ratios in the original table, and to observe the coefficient structure imposed by the model. In Table 7 groups of cells have the same estimated ratios. Examining the residuals for these groups gives all the data information along with the lack of fit at each stage of the PCC sequence. It therefore provides a valuable summary of the process. An alternative way of examining these changes is to use odds ratios, but it is omitted here to conserve space.

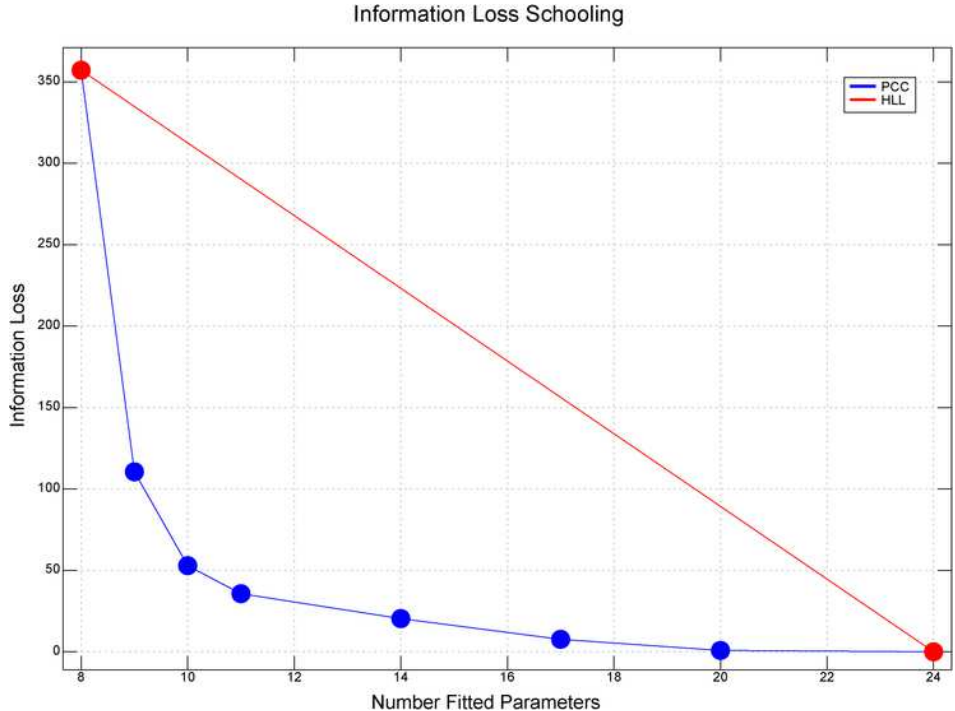

FIG. 1. *Information loss by number of model parameters for the schooling and age group data discussed by Wermuth and Cox.*

TABLE 6
*Summary tables corresponding to rows of Table 4 and Pearson ratios for the current summary*

| | (a) Current table | | | | | (b) Pearson ratios | | | | |
|---|---|---|---|---|---|---|---|---|---|---|
| $r$ | | | | | | | | | | |
| 0 | 12 | 13 | 12 | 20 | 7 | 0.873 | 0.657 | 0.814 | 1.652 | 1.941 |
| | 215 | 507 | 493 | 460 | 137 | 0.552 | 0.905 | 1.181 | 1.342 | 1.342 |
| | 277 | 300 | 192 | 126 | 38 | 1.382 | 1.040 | 0.893 | 0.714 | 0.723 |
| | 52 | 91 | 47 | 15 | 6 | 1.147 | 1.394 | 0.967 | 0.376 | 0.505 |
| | 233 | 225 | 102 | 74 | 19 | 1.661 | 1.114 | 0.678 | 0.599 | 0.516 |
| 1 | 12 | 13 | 12 | 27 | | 0.873 | 0.657 | 0.814 | 1.718 | |
| | 215 | 507 | 493 | 597 | | 0.552 | 0.905 | 1.181 | 1.342 | |
| | 277 | 300 | 192 | 164 | | 1.382 | 1.040 | 0.893 | 0.716 | |
| | 52 | 91 | 47 | 21 | | 1.147 | 1.394 | 0.967 | 0.405 | |
| | 233 | 225 | 102 | 93 | | 1.661 | 1.114 | 0.678 | 0.580 | |
| 2 | 227 | 520 | 505 | 624 | | 0.563 | 0.896 | 1.169 | 1.354 | |
| | 277 | 300 | 192 | 164 | | 1.382 | 1.040 | 0.893 | 0.716 | |
| | 52 | 91 | 47 | 21 | | 1.147 | 1.394 | 0.967 | 0.405 | |
| | 233 | 225 | 102 | 93 | | 1.661 | 1.114 | 0.678 | 0.580 | |
| 3 | 227 | 520 | 505 | 624 | | 0.563 | 0.896 | 1.169 | 1.354 | |
| | 510 | 525 | 294 | 257 | | 1.497 | 1.070 | 0.805 | 0.660 | |
| | 52 | 91 | 47 | 21 | | 1.147 | 1.394 | 0.967 | 0.405 | |
| 4 | 227 | 520 | 505 | 624 | | 0.563 | 0.896 | 1.169 | 1.354 | |
| | 562 | 616 | 341 | 278 | | 1.456 | 1.108 | 0.824 | 0.630 | |
| 5 | 227 | 520 | 1129 | | | 0.563 | 0.896 | 1.265 | | |
| | 562 | 616 | 619 | | | 1.456 | 1.108 | 0.724 | | |
| 6 | 747 | 1129 | | | | 0.760 | 1.265 | | | |
| | 1178 | 619 | | | | 1.251 | 0.724 | | | |
| 7 | 1925 | 1748 | | | | 1.000 | 1.000 | | | |
| 8 | 1925 | 1748 | | | | 1.000 | 1.000 | | | |

By row 4 of the fitting process, the key for dimension 0 is 0 0 1 1 1 and the key for dimension 1 is 0 1 2 3 3. These specify the way in which the rows and columns of the model from row 4 in Table 6 are expanded to generate the model of the initial table.

4.3. *The Christensen abortion opinion table.* The procedure we have outlined generalizes to higher dimensions. Christensen (1997) uses the data in Table 8 on opinions about abortion given without further information on its source. Table 9 is an example of the new information loss table. We still examine all pairwise comparisons for variable 3, but now we display the information loss for the difference in the joint distribution across all of the remaining variables. Each term has 11 degrees of freedom and the terms above the diagonal show the similarity of adjacent groups. Table 10 summarizes the steps and shows in row 1 that we can aggregate the two highest age categories, in row 2 aggregate the middle two age categories, and in row 3 aggregate the lowest two age categories, all with no loss of information as shown by values from the dev (term) column. Row 4 shows that collapsing over the sex categories would not be accepted using a conventional 5% significance level for $G^2$ but differential patterns of response by sex are not very large for these data, and the table can perhaps be reduced to a $2 \times 3 \times 2$ table for race, opinion and age. The resulting table is a third of the size of the original one. Row 5 shows that any further steps lead to larger noncentrality terms and significant loss of information.

Table 7
*Ratios of fitted values to independence model for each row of Table 6*

| | | | | | |
|---|---|---|---|---|---|
| 0 | 0.87 | 0.66 | 0.81 | 1.65 | 1.94 |
|   | 0.55 | 0.90 | 1.18 | 1.34 | 1.34 |
|   | 1.38 | 1.04 | 0.89 | 0.71 | 0.72 |
|   | 1.15 | 1.39 | 0.97 | 0.38 | 0.50 |
|   | 1.66 | 1.11 | 0.68 | 0.60 | 0.52 |
| 1 | 0.87 | 0.66 | 0.81 | 1.72 | 1.72 |
|   | 0.55 | 0.90 | 1.18 | 1.34 | 1.34 |
|   | 1.38 | 1.04 | 0.89 | 0.72 | 0.72 |
|   | 1.15 | 1.39 | 0.97 | 0.41 | 0.41 |
|   | 1.66 | 1.11 | 0.68 | 0.58 | 0.58 |
| 2 | 0.56 | 0.90 | 1.17 | 1.35 | 1.35 |
|   | 0.56 | 0.90 | 1.17 | 1.35 | 1.35 |
|   | 1.38 | 1.04 | 0.89 | 0.72 | 0.72 |
|   | 1.15 | 1.39 | 0.97 | 0.41 | 0.41 |
|   | 1.66 | 1.11 | 0.68 | 0.58 | 0.58 |
| 3 | 0.56 | 0.90 | 1.17 | 1.35 | 1.35 |
|   | 0.56 | 0.90 | 1.17 | 1.35 | 1.35 |
|   | 1.50 | 1.07 | 0.80 | 0.66 | 0.66 |
|   | 1.15 | 1.39 | 0.97 | 0.41 | 0.41 |
|   | 1.50 | 1.07 | 0.80 | 0.66 | 0.66 |
| 4 | 0.56 | 0.90 | 1.17 | 1.35 | 1.35 |
|   | 0.56 | 0.90 | 1.17 | 1.35 | 1.35 |
|   | 1.46 | 1.11 | 0.82 | 0.63 | 0.63 |
|   | 1.46 | 1.11 | 0.82 | 0.63 | 0.63 |
|   | 1.46 | 1.11 | 0.82 | 0.63 | 0.63 |
| 5 | 0.56 | 0.90 | 1.26 | 1.26 | 1.26 |
|   | 0.56 | 0.90 | 1.26 | 1.26 | 1.26 |
|   | 1.46 | 1.11 | 0.72 | 0.72 | 0.72 |
|   | 1.46 | 1.11 | 0.72 | 0.72 | 0.72 |
|   | 1.46 | 1.11 | 0.72 | 0.72 | 0.72 |
| 6 | 0.76 | 0.76 | 1.26 | 1.26 | 1.26 |
|   | 0.76 | 0.76 | 1.26 | 1.26 | 1.26 |
|   | 1.25 | 1.25 | 0.72 | 0.72 | 0.72 |
|   | 1.25 | 1.25 | 0.72 | 0.72 | 0.72 |
|   | 1.25 | 1.25 | 0.72 | 0.72 | 0.72 |
| 7 | 1.00 | 1.00 | 1.00 | 1.00 | 1.00 |
|   | 1.00 | 1.00 | 1.00 | 1.00 | 1.00 |
|   | 1.00 | 1.00 | 1.00 | 1.00 | 1.00 |
|   | 1.00 | 1.00 | 1.00 | 1.00 | 1.00 |
|   | 1.00 | 1.00 | 1.00 | 1.00 | 1.00 |

TABLE 8
*Christensen's abortion opinion data*

| Race | | Sex | | | Opinion on legalised abortion | Age | | | | | |
|------|--|-----|--|--|-------------------------------|-----|--|--|--|--|--|
| | | | | | | **0**<br>18–25 | **1**<br>26–35 | **2**<br>36–45 | **3**<br>46–55 | **4**<br>56–65 | **5**<br>66+ |
| 0 | white | 0 | male | 0 | supports | 96 | 138 | 117 | 75 | 72 | 83 |
| | | | | 1 | opposes | 44 | 64 | 56 | 48 | 49 | 60 |
| | | | | 2 | undecided | 1 | 2 | 6 | 5 | 6 | 8 |
| | | 1 | female | 0 | supports | 140 | 171 | 152 | 101 | 102 | 111 |
| | | | | 1 | opposes | 43 | 65 | 58 | 51 | 58 | 67 |
| | | | | 2 | undecided | 1 | 4 | 9 | 9 | 10 | 16 |
| 1 | nonwhite | 0 | male | 0 | supports | 24 | 18 | 16 | 12 | 6 | 4 |
| | | | | 1 | opposes | 5 | 7 | 7 | 6 | 8 | 10 |
| | | | | 2 | undecided | 2 | 1 | 3 | 4 | 3 | 4 |
| | | 1 | female | 0 | supports | 21 | 25 | 20 | 17 | 14 | 13 |
| | | | | 1 | opposes | 4 | 6 | 5 | 5 | 5 | 5 |
| | | | | 2 | undecided | 1 | 2 | 1 | 1 | 1 | 1 |

For this example, the second step in table analysis, constructing a HLLM, is valuable. We fit the HLLM using backward selection based on information and information gradients rather than by the more traditional approach using statistical significance. The table has only one item in dimension one, so the corresponding log-linear model terms are zero. Table 11 shows an extract from the report on a HLLM fitted to the four way table associated with line 4. It is an example of a HLLP model. Since the distribution across the other variables is independent of the sex variable, we omit all interaction terms involving sex. We retain Sex in the model specification because the sex marginal will be preserved in the model of the entire data.

TABLE 9
*Information loss for age categories*

| | **0** | **1** | **2** | **3** | **4** | **5** |
|---|-------|-------|-------|-------|-------|-------|
| 0 | | 7.21 | 14.29 | 22.21 | 35.21 | 54.45 |
| 1 | | | 7.05 | 15.24 | 22.48 | 38.21 |
| 2 | | | | 4.58 | 9.87 | 19.60 |
| 3 | | | | | 3.43 | 9.59 |
| 4 | | | | | | 2.19 |
| 5 | | | | | | |

Table 10
*PCC steps for abortion data*

| r | d | Key | | | | | | | dim | | | | dev | dfmod | dfres | dev (term) | df | AdRsq |
|---|---|---|---|---|---|---|---|---|---|---|---|---|---|---|---|---|---|---|
| 0 |   |   |   |   |   |   |   |   | 2 | 2 | 3 | 6 | 0.00 | 71 | 0 | 0.00 | 0 | 1.000 |
| 1 | 3 | 0 | 1 | 2 | 3 | 4 | 4 |   | 2 | 2 | 3 | 5 | 2.19 | 20 | 60 | 2.19 | 11 | 0.898 |
| 2 | 3 | 0 | 1 | 2 | 2 | 3 | 3 |   | 2 | 2 | 3 | 4 | 6.77 | 17 | 49 | 4.58 | 11 | 0.843 |
| 3 | 3 | 0 | 0 | 1 | 1 | 2 | 2 |   | 2 | 2 | 3 | 3 | 13.98 | 14 | 38 | 7.21 | 11 | 0.784 |
| 4 | 1 | 0 | 0 |   |   |   |   |   | 2 | 1 | 3 | 3 | 42.65 | 11 | 21 | 28.67 | 17 | 0.565 |
| 5 | 0 | 0 | 0 |   |   |   |   |   | 1 | 1 | 3 | 3 | 65.87 | 10 | 13 | 23.21 | 8 | 0.420 |
| 6 | 3 | 0 | 0 | 1 | 1 | 1 | 1 |   | 1 | 1 | 3 | 2 | 77.61 | 9 | 11 | 11.74 | 2 | 0.340 |
| 7 | 2 | 0 | 0 | 1 |   |   |   |   | 1 | 1 | 2 | 2 | 93.28 | 8 | 10 | 15.67 | 1 | 0.219 |
| 8 | 2 | 0 | 0 | 0 |   |   |   |   | 1 | 1 | 1 | 2 | 121.47 | 8 | 9 | 28.19 | 1 | 0.000 |
| 9 | 0 | 0 | 0 |   |   |   |   |   | 1 | 1 | 1 | 2 | 121.47 | 8 | 9 | 0.00 | 1 | 0.000 |

The saturated model for the data is [roa] and the main effect [s] is included because we expand the condensed model to match the marginals of the table, assuming independence of categories within the partitions. For this reduced model, given the marginals of each of the variables, the two terms of the [oa] interaction and a single parameter for an [or] interaction are associated with most of the information in the table. Christensen reaches a similar conclusion—that these two interactions provide a nearly complete summary of the data—but after a much more extensive discussion. We have shown that a model of the same structure but with fewer categories contains nearly all the information.

Table 11
*HLLP model for line 4 from Table 10*

| r | Terms | dev | dfmod | dfres | dev (term) | df | AdRsq |
|---|---|---|---|---|---|---|---|
| 0 | rsoa | 0.000 | 17 | 0 | 0.000 | 0 | 1.000 |
| 7 | roa s | 0.000 | 17 | 0 | 0.000 | 0 | — |
| 8 | oa ra ro s | 5.245 | 13 | 4 | 5.245 | 4 | 0.800 |
| 9 | oa ro s | 9.225 | 11 | 6 | 3.980 | 2 | 0.766 |
| 10 | oa s r | 23.214 | 9 | 8 | 13.989 | 2 | 0.558 |
| 11 | a o s r | 78.811 | 5 | 12 | 55.597 | 4 | 0.000 |

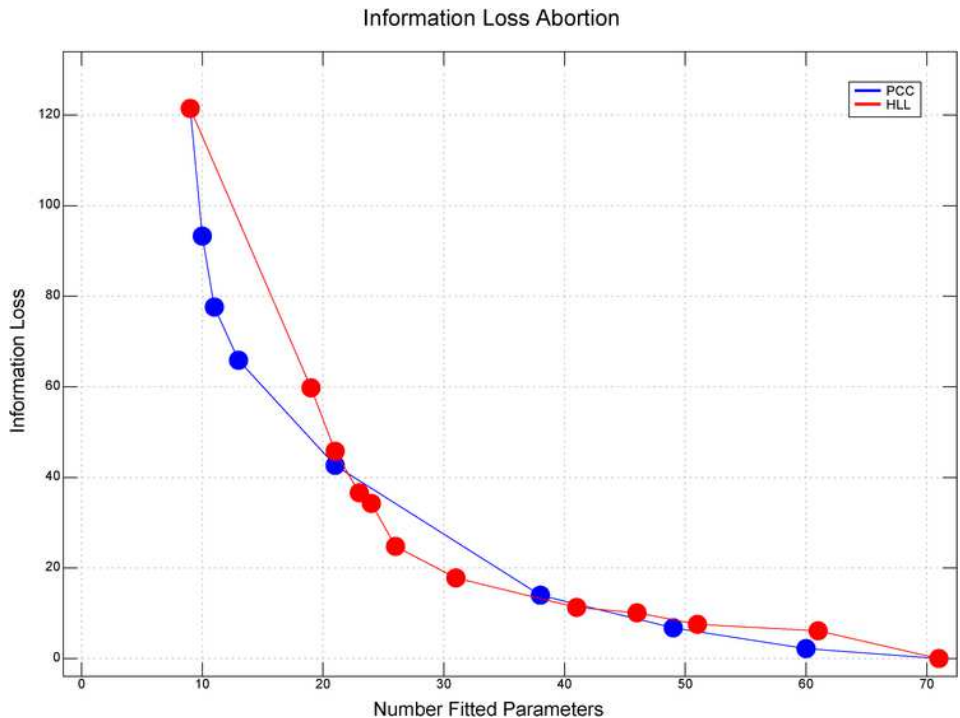

Fig. 2.  *Information loss alternative procedures with abortion opinion data.*

Figure 2 shows in this case a PCC sequence provides a very similar relationship between the information loss and number of parameters to the sequence using a HLL model. There is a sense in which they provide approximately equally informative summaries for a given number of parameters.

Most examples of HLLMs in the statistical literature are based on examples with few category levels. If there are only two, our collapsing procedure is equivalent to collapsing over the variable. This is not necessarily helpful with such models and is not really designed to deal with tables of that structure. Even so, in nearly all cases we have examined in which some variables have more than two categories the PCC model gives much lower information loss statistics than an HLLM in the region near the model of marginal independence for all variables.

**5. Large census tables.**   Working with and developing parsimonious models for large data sets is much easier using PCC since the procedure permits rapid scanning of large tables. In the case of nominal categories, the calculation time expands as the square of the number of categories, so tables with a few variables but large numbers of categories are less tractable than tables with more variables and smaller sets of categories for each variable.

In official collections there are often between 10 and 20 categories associated with a variable, but the calculations remain feasible. Variables with larger numbers of cases are discussed below.

We have used PCC for tables with 7 variables, 50 million cells, 600,000 cells with nonzero counts, and up to 30 categories per variable with interesting and useful results that we will report elsewhere. To handle such large contingency tables requires the use of sparse array methods. The procedure may remain feasible for much larger tables.

Census data often yield far bigger contingency tables and the decisions on how to summarize and report the data are inevitably complex and almost always ad hoc. For example, the 2000 U.S. census long form, completed by a sample of one in six households nationwide, contains 53 questionnaire items, most of which are categorical, and these data are, in principle, available for over 3000 counties and even lower levels of geography in many instances. Many of the categorical variables have very large numbers of categories, some with ordinal structure and some not. The U.S. Census Bureau has developed an elaborate confidentiality protection scheme based on the reporting of a large number of marginal tables from the data and made only these margins available.[1]

To illustrate the application to more reasonably-sized tables, we explore the analysis of a table with 9,680 cells drawn from a slightly altered version of the 1981 Australian population census, involving about 10 million individuals.[2]

We consider a table where individuals are cross-classified by Age (11) × Marital Status (5) × Qualification (11) × Family Income (16), yielding a table with 9,680 cells. The numbers are so large that the information associated with higher order interactions is large. Only a saturated model really fits this data. Figure 3 shows that by using the partitioning model and a linear scale we can associate a large part of the information with a modest number of parameters in a collapsed table.

Over the entire region down to about 500 parameters the PCC model is clearly dominated by the HLLM. Retaining the ability of the HLLM to have arbitrary patterns of the odds ratios within each of the factor interaction terms permits closer modeling of the cell values, but having regard to the huge numbers of observations, one needs to think very carefully about the accuracy and adequacy of the data generating process in giving weight to these differences.

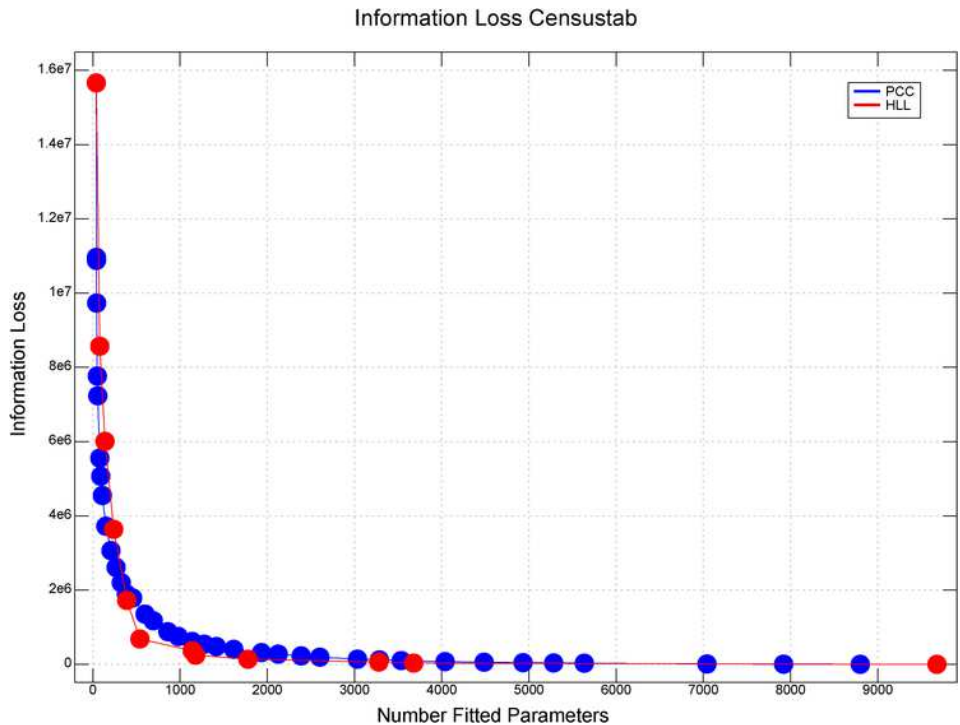

FIG. 3. *Information Loss with Census table (Age × Marital Status × Qualification × Family Income.*

Table 12 provides information which helps us examine effects of collapsing this table. It is not a full listing of the PCC output, but gives the first six lines and then every fifth line, and omits the aggregation keys to save space. There remains sufficient information to see the main character of what is happening.

Dimension 2 distinguishes qualifications. Aggregation of two of the qualification categories leads to no loss of information in row 1. Thereafter, there are no subtables where the category on some dimension is independent of the distribution of other variables. Row 2 shows that the least of these has a noncentrality parameter about 1.6. Each further step leads to some further loss, but by line ($r = 10$) it has not reached 1% and by line 15 has reached only about 2% of the deviance from the independence model. A table one fifth the size expanded to the original margins gives 98% of the information about associations and one tenth size only drops the information to 95% of the original table.

For this table there are 15 terms in a saturated HLLM. For tables of census data available to us with up to four variables the data reject any model short of the saturated model, using as a very crude measure the $G^2$ statistic calculated for all nonzero cells, with no adjustments for the zeros made to the degrees of freedom. See Fienberg and Rinaldo (2007) for a discussion of such adjustments. Even with these assumptions the noncentrality parameter must be several multiples of the number of degrees of freedom. At this point in the sequence of collapsing steps, the key vectors reflected the ordinal structure of the data except for the final variable, Family Income. Although the ordinal pattern is clearly a feature of this variable, it showed some differences from the ordinal structure with the key: 0 1 1 1 0 2 3 3 1 1 4 4 5 5 6 7. This is not surprising as Family Income is an aggregate variable with a complex interaction between family circumstances and income sources, so requiring ordinality for this particular variable may not be appropriate. This substantial simplification at various points in the PCC sequence is clearly helpful in constructing simpler descriptions.

TABLE 12
*Lines from the PCC summary for the Australian census data extract*

| r | d | | dim | | | dev | dfmod | dfres | dev (term) | df | AdRsq |
|---|---|---|---|---|---|---|---|---|---|---|---|
| 0 | | 11 | 5 | 11 | 16 | 0.00 | 9679 | 0 | 0.00 | 0 | 1.000 |
| 1 | 2 | 11 | 5 | 10 | 16 | 725.04 | 8800 | 879 | 725.04 | 879 | 0.999 |
| 2 | 2 | 11 | 5 | 9 | 16 | 4515.16 | 7921 | 1758 | 3790.12 | 879 | 0.998 |
| 3 | 2 | 11 | 5 | 8 | 16 | 10226.25 | 7042 | 2637 | 5711.09 | 879 | 0.998 |
| 4 | 1 | 11 | 4 | 8 | 16 | 31271.80 | 5635 | 4044 | 21045.56 | 1407 | 0.995 |
| 5 | 3 | 11 | 4 | 8 | 15 | 38089.68 | 5284 | 4395 | 6817.88 | 351 | 0.995 |
| 10 | 3 | 9 | 4 | 7 | 13 | 119605.04 | 3285 | 6394 | 17616.83 | 251 | 0.988 |
| 15 | 3 | 8 | 4 | 6 | 10 | 316797.86 | 1934 | 7745 | 41883.01 | 191 | 0.975 |
| 20 | 0 | 6 | 4 | 5 | 8 | 750387.67 | 979 | 8700 | 133362.68 | 159 | 0.947 |
| 25 | 3 | 6 | 4 | 3 | 5 | 1892492.24 | 384 | 9295 | 104126.96 | 71 | 0.875 |
| 30 | 3 | 5 | 4 | 2 | 2 | 4549042.79 | 109 | 9570 | 826921.73 | 39 | 0.707 |
| 35 | 3 | 3 | 3 | 1 | 1 | 9731662.12 | 43 | 9636 | 1965711.15 | 8 | 0.378 |
| 39 | 0 | 1 | 2 | 1 | 1 | 15663099.18 | 39 | 9640 | 0.00 | 1 | 0.000 |

It is interesting to examine the HLLP models selected using the PCC sequence. Table 13(a) gives the analysis for the original data, and Table 13(b) the HLLP model for the line 20 table with 960 cells. The table shows that the PCC collapsing steps which have reduced the number of cells from 9,680 to 960 have made no difference at all to the sequence of steps followed in the HLL calculations, and none of the adjusted $R^2$ values at any stage of the process differ by more than 1 percentage point. The collapsed table shows all the main features of the source data. Table 13(c) gives a much more radical collapse and over a third of the total information loss. We have collapsed this table to less than 0.5% of the source table size, but much of the structure is still preserved. In this case there are differences in the sequence in the HLLP model, but several points in the sequence are still the same. Estimates of the noncentrality parameters are broadly preserved. There is no change in the set of significant terms using conventional criteria. In other cases we have examined, there are examples where there are some differences in the sequence of steps selected, but they have been for only a few steps until the original sequence is re-established.

Clearly, going too far in the process will lose sufficient information to cause difficulties, but an information loss of 5 to 10% often leads to virtually no change in the HLL modeling, with a huge gain in the ability to describe the data in simple ways.

We have been careful to distinguish the PCC process within the class of HLLP models. Alternative procedures for determining the set of partitions can be used, but the PCC sequence has a number of simple and attractive features.

**6. Further observations.** We can learn a great deal about the PCC process from examining a wide range of examples. Jackson, Gray and Fienberg (2007) we provide examples from the literature and some detailed comparisons of analyses from several multi-dimensional tables using the PCC framework and our original analyses. Here we comment on some aspects from that experience.

The PCC process may not generate a set of categories which provides the minimal information loss across all possible category groupings. The limitations of single step procedures in seeking clusters in multi-dimensional problems are well known. A minimal step at some stage may preclude other alternatives at a later stage and prevent the sequence from passing through some minimal points. Analysis of the Wermuth and Cox data set used by Dellaportas and Tarantola showed that neither the PCC procedure nor their Bayesian model selection procedure found the global minimal information loss solution.

For small models it is possible to examine all possible partitions of the categories and evaluate the information loss associated with those partitions. For small numbers of categories, it is possible to consider all partitions across each dimension of a table, but as the Bell Numbers $1, 2, 5, 15, 52, 203, 877, \ldots$ grow rapidly it quickly becomes infeasible. For a $4 \times 4 \times 4 \times 4$ table there are $15^4$ options, so it can be explored, but anything much larger can quickly become infeasible. It is clearly not feasible for most of the tables of interest for large data sets with moderate numbers of categories per variable.

In some cases, where there are a group of categories with a small distance between them, we could also consider a search procedure which takes one or two further steps before making a decision. Our experience so far, however, is that the simple local valley based on following our procedure provides useful summaries, a valuable guide to the information structure, and improved confidence in the model structure obtained.

TABLE 13
*Hierarchical log-linear models at PCC steps for the Australian census data extract*

| r | Terms | dev | dfmod | dfres | dev (term) | df | AdRsq |
|---|-------|-----|-------|-------|------------|-----|-------|
| | | (a) Source table—dimensions 11 5 11 16 | | | | | |
| 0 | AMQF | 0.000 | 9679 | 0 | 0.000 | 0 | 1.000 |
| 1 | MQF AQF AMF AMQ | 33887.855 | 3679 | 6000 | 33887.855 | 6000 | 0.997 |
| 2 | MQF AQF AMF | 61348.366 | 3279 | 6400 | 27460.512 | 400 | 0.994 |
| 3 | MQF AMF AQ | 136770.866 | 1779 | 7900 | 75422.500 | 1500 | 0.989 |
| 4 | QF AMF MQ AQ | 247723.904 | 1179 | 8500 | 110953.038 | 600 | 0.982 |
| 5 | QF AMF AQ | 365642.032 | 1139 | 8540 | 117918.128 | 40 | 0.974 |
| 6 | QF MF AF AQ AM | 682961.719 | 539 | 9140 | 317319.687 | 600 | 0.954 |
| 7 | MF AF AQ AM | 1721725.583 | 389 | 9290 | 1038763.863 | 150 | 0.886 |
| 8 | MF AQ AM | 3636440.284 | 239 | 9440 | 1914714.702 | 150 | 0.763 |
| 9 | MF Q AM | 6006085.780 | 139 | 9540 | 2369645.496 | 100 | 0.613 |
| 10 | F Q AM | 8569234.731 | 79 | 9600 | 2563148.951 | 60 | 0.451 |
| 11 | F Q M A | 15663099.176 | 39 | 9640 | 7093864.445 | 40 | 0.000 |
| | | (b) Condensed table from line 20—dimensions 6 4 5 8 | | | | | |
| 0 | AMQF | 0.000 | 959 | 0 | 0.000 | 0 | 1.000 |
| 1 | MQF AQF AMF AMQ | 20254.229 | 539 | 420 | 20254.229 | 420 | 0.997 |
| 2 | MQF AQF AMF | 36045.634 | 479 | 480 | 15791.405 | 60 | 0.995 |
| 3 | MQF AMF AQ | 78960.937 | 339 | 620 | 42915.303 | 140 | 0.992 |
| 4 | QF AMF MQ AQ | 160606.450 | 255 | 704 | 81645.513 | 84 | 0.986 |
| 5 | QF AMF AQ | 271689.641 | 243 | 716 | 111083.191 | 12 | 0.976 |
| 6 | QF MF AF AQ AM | 529426.958 | 138 | 821 | 257737.317 | 105 | 0.959 |
| 7 | MF AF AQ AM | 1466497.479 | 110 | 849 | 937070.521 | 28 | 0.891 |
| 8 | MF AQ AM | 3192037.053 | 75 | 884 | 1725539.574 | 35 | 0.772 |
| 9 | MF Q AM | 5455415.358 | 55 | 904 | 2263378.305 | 20 | 0.620 |
| 10 | F Q AM | 7934859.751 | 34 | 925 | 2479444.393 | 21 | 0.459 |
| 11 | F Q M A | 14912711.502 | 19 | 940 | 6977851.752 | 15 | 0.000 |
| | | (c) Condensed table from line 32—dimensions 4 3 2 2 | | | | | |
| 0 | AMQF | 0.000 | 47 | 0 | 0.000 | 0 | 1.000 |
| 1 | MQF AQF AMF AMQ | 51.212 | 41 | 6 | 51.212 | 6 | 1.000 |
| 2 | AQF AMF AMQ | 186.612 | 39 | 8 | 135.400 | 2 | 1.000 |
| 3 | QF AMF AMQ | 1702.175 | 36 | 11 | 1515.564 | 3 | 0.999 |
| 4 | QF AMF MQ AQ | 5157.652 | 30 | 17 | 3455.477 | 6 | 0.999 |
| 5 | QF AMF AQ | 25974.274 | 28 | 19 | 20816.622 | 2 | 0.995 |
| 6 | AMF AQ | 48953.668 | 27 | 20 | 22979.394 | 1 | 0.990 |
| 7 | MF AF AQ AM | 85517.442 | 21 | 26 | 36563.774 | 6 | 0.987 |
| 8 | AF AQ AM | 832835.299 | 19 | 28 | 747317.857 | 2 | 0.882 |
| 9 | F AQ AM | 2057950.185 | 16 | 31 | 1225114.886 | 3 | 0.737 |
| 10 | F Q AM | 3687777.788 | 13 | 34 | 1629827.603 | 3 | 0.571 |
| 11 | F Q M A | 10107312.737 | 7 | 40 | 6419534.948 | 6 | 0.000 |

An interesting and more general alternative to the scheme outlined so far is to permit different partitions in different terms within a hierarchic log-linear model. Such a model would not be a hierarchic model on partitions of the categories. An [AB] term might be fitted better with a different partition of the categories of A from an [AC] term. The procedure explored in Sections 2 and 3 can be applied to any table to clarify its information structure.

Because categories with smaller numbers of cases have less information, the PCC commonly aggregates small categories with a much larger one, or with another small category. The number of cells which are problematic can reduce rapidly with this aggregation, but even in condensed tables confidentiality problems can arise. These issues and a range of other examples are examined in Jackson, Gray and Fienberg (2007).

6.1. *Where is the information in a contingency table?* The deviance vs parameter graph is very informative about the structure of a contingency table. If it is nearly a straight line, then either the table is nearly all noise, or the information is widely dispersed and all collapsing steps generate similar average significant information loss. If you consider the Gini like coefficient for the graph, based on the ratio of the area under the curve to the area of the triangle, very small values indicate a high degree of concentration of the information in a manner associated with the category structure. The tools in this paper give two ways of exploring that. The PCC framework helps clarify the coefficient structure for individual components of an HLLM. It generates a sequence of models within the HLLP class and the simplification associated with the collapsed data often helps explain and characterize the structure associated with the terms in the HLLM. The PCC framework provides a consistent way of thinking about increasingly detailed models, with a shifting boundary of the effects considered as *noise* depending on the threshold necessary for a particular analysis. However, other processes can also generate sequences of HLLP models and we see PCC as a first step in exploring them. For all of them the gradient of the information curve will provide insight into the magnitude of the effects being ignored at any level of description.

6.2. *Guided category collapse: priors or data based.* In Table 2 we gave the deviance associated with alternative category aggregates for schooling. It shows clearly the magnitude of the effect of different category aggregates on the deviance from the independence model. The impact of aggregating two categories can rank from none (the deviance is of the order expected from independent sampling alone) to approximately two-thirds of the total deviance from independence. Even considering adjacent pairs, there are alternative ways of grouping the categories, and an inappropriate choice based solely on the user's prior views may have a high risk of generating a poor outcome. In some large tables, three or four inappropriate aggregations can largely destroy the information about variable interactions in a table. Even in small tables such as the one in Tables 1 and 3 showed that aggregating age categories 0 and 3 would be very destructive of the information about the association between them. That is an unlikely pair to be selected, but it does provide a warning of the potential for large loss when information loss tables are not used to guide selection. It highlights the difficulty of meta analysis and comparing analyses where different categories or concepts are used. Differences in category combinations can have large effects. We strongly recommend at least construction of an information loss table, as illustrated in Tables 2, 3 and 10, before any category aggregation is undertaken prior to data publication and some reporting of its effects when it is used.

**Acknowledgments.** We are grateful for discussion with participants in a Statistics New Zealand seminar and for valuable comments from Nanny Wermuth, Alan Lee and Stephen Haslett, as well as from an Associate Editor and two referees, whose comments led to substantial improvements in the presentation.

## SUPPLEMENTARY MATERIAL

**Tools for construction and comparison of PCC and HLL models** (DOI: 10.1214/08-AOAS175SUPP; .zip). A program to execute the procedures in the paper are provided in the supplementary material and illustrated with steps to generate some graphs and tables from the paper. The read.me file provides some instructions on its use. The program requires the free array programming language J available from **http://www.jsoftware.com**.

L. F. JACKSON
EMERITUS PROFESSOR
SCHOOL OF ECONOMICS AND FINANCE
VICTORIA UNIVERSITY OF WELLINGTON
WELLINGTON
NEW ZEALAND
E-MAIL: fraser.jackson@vuw.ac.nz

A. G. GRAY
STATISTICAL RESEARCH ASSOCIATES
WELLINGTON
NEW ZEALAND
E-MAIL: alistair@statsresearch.co.nz

S. E. FIENBERG
DEPARTMENT OF STATISTICS
    AND MACHINE LEARNING DEPARTMENT
CARNEGIE MELLON UNIVERSITY
PITTSBURGH, PENNSYLVANIA 15213-3890
USA
E-MAIL: fienberg@stat.cmu.edu